\documentclass[twocolumn, pre, 10pt, showpacs, english, preprintnumbers, amsmath, amssymb, superscriptaddress, aps,longbibliography]{revtex4-2}

\usepackage{csquotes}
\usepackage{comment}
\usepackage{graphicx}
\usepackage[dvipsnames]{xcolor}
\usepackage[normalem]{ulem}
\usepackage{appendix}
\usepackage{silence}
\renewcommand\vec{\boldsymbol}

\definecolor{orange}{rgb}{1,0.5,0}
\definecolor{goodgreen}{rgb}{0.1,0.5,0}
\definecolor{goodred}{rgb}{0.7,0,0}
\usepackage{lineno}

\usepackage[colorlinks,urlcolor=goodgreen,citecolor=blue,linkcolor=goodred]{hyperref}

\usepackage{float}
\graphicspath{ {figures/} }

\usepackage{babel}
\begin{document}

\title{Spin-Splitter and Inverse Effects in Altermagnetic Hybrid Structures}
\newcommand{\orcid}[1]{\href{https://orcid.org/#1}{\includegraphics[width=8pt]{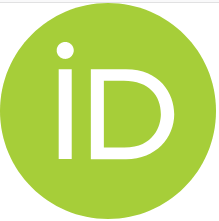}}}

\author{Nicolas Sigales \orcid{0009-0008-2302-4239}}
\email{nsigales@fagro.edu.uy}
\affiliation{Instituto de Física, Facultad de Ciencias, UdelaR, Montevideo, Uruguay}

\author{Tim Kokkeler\orcid{0000-0001-8681-3376}}
\email{tim.h.kokkeler@jyu.fi}
\affiliation{Department of Physics and Nanoscience Center, University of Jyvaskyla,
P.O. Box 35 (YFL), FI-40014 University of Jyvaskyla, Finland}

\author{Gonzalo De Polsi\orcid{0000-0001-5741-4560}}
\affiliation{Instituto de Física, Facultad de Ciencias, UdelaR, Montevideo, Uruguay}

\author{Sebastian Bergeret\orcid{0000-0001-6007-4878}}
\affiliation{Centro de Física de Materiales (CFM-MPC) Centro Mixto CSIC-UPV/EHU, E-20018 Donostia-San Sebastián,  Spain}
\affiliation{Donostia International Physics Center (DIPC), 20018 Donostia-San Sebastián, Spain}

\begin{abstract}
We 
provide a theoretical 
description of diffusive
charge and spin transport in hybrid devices containing altermagnets. Based on recently derived drift–diffusion equations for coupled charge and spin dynamics and general boundary conditions, our approach provides a unified description of the spin-splitter effect, i.e., the conversion of charge currents into spin currents, and its inverse in terms of experimentally accessible parameters. We analyze, analytically and numerically, the spin-splitter effect, demonstrating that an injected spin accumulation generates a measurable 
voltage difference across the transverse direction in the altermagnet. Motivated by a recent experiment, we also analyze a nonlocal spin-valve geometry in which an altermagnetic strip injects spin into a diffusive normal metal. We derive the resulting nonlocal voltage detected by a ferromagnetic electrode as a function of the relative orientation of the Néel vector and the ferromagnetic polarization, accounting for the main experimental findings. For this setup, we further address spin precession during diffusive transport by analyzing the spin Hanle effect. Our results provide theoretical explanations and predictions for several altermagnet hybrid structures
\end{abstract}
\maketitle

\section{Introduction}
Since its prediction \cite{sinova2022emerging}, altermagnetism 
has attracted a great deal of attention in the condensed-matter community, both theoretically \cite{gonzalez2021efficient,das2023transport,zhang2024electric,gomonay2024structure,maeda2024theory,sukhachov2024impurity,sivianes2025optical,yang2024symmetry,li2023majorana,li2024realizing,zarzuela2025transport,he2025evidence}, experimentally \cite{reimers2024direct,fedchenko2024observation,guo2024direct, chen2025spin,li2025exploration,liu2024giant}, and in connection with superconductivity \cite{lu2024phi,Kokkeler2025,heras2025interplay,mondal2024distinguishing,leiviska2024anisotropy,cheng2024field,fukaya2025superconducting,sun2023andreev,giil2024superconductor,Debnath2025}. For further references, we refer to the reviews in Refs.~\cite{jungwirth2024altermagnets,bai2024altermagnetism,jungwirth2025altermagnetic} and the references therein.
Altermagnets (AM) are characterized by a vanishing net magnetization, as in antiferromagnets, while exhibiting a nonrelativistic spin splitting of their electronic bands, as in ferromagnets.
This unique combination of properties defines a distinct class of magnetic materials that bridges the conventional distinction between ferromagnetic and antiferromagnetic order.

Beyond their fundamental interest, AM have recently emerged as a promising platform for spintronics applications, enabling the efficient generation and manipulation of spin currents without net magnetization or stray fields.
In particular, the transport properties of conducting AM provide valuable insight into their microscopic magnetic structure and underlying symmetry characteristics.
It was predicted in Ref.~\cite{gonzalez2021efficient} that a charge current flowing through an altermagnet can generate a spin current, in a manner analogous to the spin Hall effect, but originating from the intrinsic nonrelativistic spin splitting of the altermagnetic state rather than from spin--orbit coupling.
This phenomenon, known as the spin-splitter effect (SSE), has been supported by recent experimental observations in altermagnetic materials \cite{guo2024direct,chen2025spin}.

With the exception of a few works (see, e.g., \cite{Kokkeler2025, Linder1, Linder2}), most theoretical studies of the SSE have focused on bulk altermagnetic materials. However,  it is important to understand how this effect manifests in hybrid nanostructures, where either spin or charge currents are injected and the corresponding converted signals are detected---a standard strategy in spintronics.
Examples include nonlocal spin-valve (NLSV) geometries, commonly used to measure spin currents and spin-relaxation lengths, as well as spin--orbit-based systems, where spin-to-charge conversion is detected electrically.
More recently, an NLSV geometry employing an AM as a spin injector has been used with the aim to directly demonstrate the time-reversal-symmetry-breaking nature of spin injection in these materials \cite{mencos2025direct}.
Such transport experiments in realistic materials, \textit{i.e.} in the presence of disorder, are most accurately described within a kinetic-equation-based approach.


In this work, we provide such description for several setups and hybrid structures. Our analysis builds upon the drift--diffusion equations derived in Ref.~\cite{Kokkeler2025}. 
We analyze the spin-splitter effect, extending the results of Ref. \cite{Kokkeler2025} for contacts with finite transmission,  as well as its inverse counterpart, computing the charge current or voltage induced in an altermagnet when a spin is injected.
We demonstrate that, in a finite altermagnetic strip, the voltage difference induced across its edges is proportional to the total spin current injected into it. 
Moreover, we analyze the role of finite-size effects, interface transparency, and spin relaxation, and establish the connection between the inverse spin-splitter response and nonlocal spin-transport signals measured in multiterminal geometries.

The article is setup as follows. First, in Sec. \ref{sec:Framework} we discuss the theoretical framework that we use to consider transport in altermagnet hybrid structures. Then, in Sec. \ref{sec:SSE} we use these equations to provide a description of the spin-splitter effect in realistic systems, including disorder effects and a finite interface resistance. In Sec. \ref{sec:ISSE} we present the inverse of this effect, in which a transverse voltage can be generated via spin injection. Subsequently, in Sec. \ref{sec:hybrid} we consider the setup experimentally investigated in \cite{mencos2025direct} and provide expressions for the non-local voltage that arises in this junction. We conclude our article in Sec. \ref{sec:conclusions} with a summary of the main results and the conclusions we draw from them.

\section{Generalized Kinetic Framework for Charge and Spin Transport}\label{sec:Framework}

In this section, we briefly present  the generalized kinetic framework describing the coupled transport of charge and spin in disordered systems and altermagnets. It is based on the diffusive formalism developed for collinear altermagnets in Ref. ~\cite{Kokkeler2025}.

In dirty materials the  electric potential $\mu$ and spin chemical potentials $\mu^s$ satisfy the following diffusion equations
\begin{align}
  2\nu_0\partial_{t}\mu + \partial_{k}j_{k} &= 0\;,\label{eq:chargediffnormalAM}\\
  2\nu_0\partial_{t}\mu_a^s+ \partial_{k}j^{s}_{ka} &= -2\nu_0\Gamma_{ab}\mu_s^b\;,
\end{align}
where $\nu_0$ is the density of states per spin, which in altermagnets is guaranteed to be independent of spin, and $\Gamma_{ab}$ is the spin–relaxation tensor \cite{zhang2019theory}. In an altermagnet the spin-relaxation tensor is given by $\Gamma_{ab} = \frac{1}{\tau_s}N_aN_b + \frac{1}{\tau_{s\perp}}(\delta_{ab}-N_aN_b)$, where $\tau_s$ is the relaxation time for spin along the N\'eel vector $\vec{N}$ of the altermagnet and $\tau_{s\perp}$ is the relaxation time for spins perpendicular to the N\'eel vector.  
These equations describe the conservation of charge and spin densities in the diffusive regime, generalizing the framework to arbitrary dimensionality and relaxation anisotropy while allowing for a single, well-defined spin quantization axis.

The corresponding constitutive relations defining the charge, $\vec{j}$, and spin, $\vec{j}^s$ currents in altermagnets are
\begin{align}
  j_{k}&= -\sigma_{D} \partial_{k}\mu-\sigma_{D} N_{a}T_{jk}\partial_{j}\mu^{s}_{a}\label{eq:ElectricalCurrentNormalAlter}\;,\\
  j^{s}_{ka}&=-\sigma_{D} \partial_{k}\mu^{s}_{a}-\sigma_{D} N_{a}T_{jk}\partial_{j}\mu-\sigma_{D} N_{b}K_{jk}\epsilon_{abc}\partial_{j}\mu^{s}_{c}\;.\label{eq:SpinCurrentNormalAlter}
\end{align}
with $\sigma_D=2D\nu_0$ the Drude conductivity, $D$ the diffusion coefficient, and $\nu_0$ the density of states per spin at the Fermi level.  

The tensors $T_{jk}$ and $K_{jk}$ encode the spin--momentum coupling intrinsic to altermagnets:
$T_{jk}$ represents the linear coupling between charge and spin gradients allowed by the underlying crystal symmetry,
whereas $K_{jk}$ describes the interconversion between spin currents with spin polarizations perpendicular to the N\'eel vector,
analogous to the spin-swapping effect \cite{lifshits2009swapping} in systems with spin--orbit coupling.

The general boundary conditions that complement Eqs.~(\ref{eq:chargediffnormalAM})–(\ref{eq:SpinCurrentNormalAlter}) and govern transport across hybrid interfaces take the compact 
form 

\begin{equation}
    \begin{pmatrix}
        \hat{n}\!\cdot\!j_e \\[4pt]
        \hat{n}\!\cdot\!j_a^s
    \end{pmatrix}
    =
    -G_B
    \begin{pmatrix}
        \mu_{\mathrm{res}} - \mu \\[4pt]
        m_a\mu_{\mathrm{res}}^s -  \mu_a^s
    \end{pmatrix}-P G_B\begin{pmatrix}
\mu^s_{\mathrm{res}} - \vec{\mu}^s\cdot\vec{m} \\[4pt]
        m_a(\mu_{\mathrm{res}} -  \mu)
    \end{pmatrix}\;,
    \label{eq:KLESS}
\end{equation}
where $G_B = G_\uparrow + G_\downarrow$ is the total barrier conductance per unit area, and 
$P = ({G_\uparrow - G_\downarrow})/({G_\uparrow + G_\downarrow})$. 
Here, $\uparrow$ and $\downarrow$ refer to the spin polarization of the barrier itself, 
or to the spin quantization axis, $\vec{m}$ if a ferromagnet is one of the junction electrodes.
Eqs.~(\ref{eq:ElectricalCurrentNormalAlter})-(\ref{eq:KLESS}) establish the main equations for analyzing spin–charge conversion phenomena, such as the spin-splitter effect (SSE) and its inverse, discussed in the following sections.  

Throughout this article, we consider  collinear AMs,  in which the N\'eel vector $\vec{N}$ is homogeneous.
In this case, the spin accumulation is aligned with $\vec{N}$, and the spin chemical potential can be written as
$\mu^s_a = N_a \mu^s$.
Since the $\tau_{s\perp}$ and $K_{jk}$ terms only affect spin components perpendicular to the N\'eel vector, they do not contribute to the transport equations in this collinear regime. Thus,  we may write  a single spin relaxation time $\tau_s$.
This situation corresponds to the case relevant for monodomain $d$--wave altermagnets. 
In this collinear regime, the spin-current tensor reduces to a single component
$j^s_k \equiv N_a j^s_{ka}$.

Thus in this case  Eqs. (\ref{eq:ElectricalCurrentNormalAlter}-\ref{eq:SpinCurrentNormalAlter}) for  the charge and spin currents reduce to
\begin{equation}
\begin{cases}
j_k = -\sigma_D\left(\partial_k\mu + T_{kj}\partial_j \mu^s\right)\;, \\[4pt]
j^s_k = -\sigma_D\left(\partial_k \mu^s + T_{jk}\partial_j\mu\right)\;.
\end{cases}
\label{eq:currents}
\end{equation}

In the remaining sections we will use the above equations and boundary conditions to describe the spin-splitter effect,  
 \ref{sec:SSE}, its inverse,  \ref{sec:ISSE}, and  transport in a non-local spin valve \ref{sec:hybrid}.   {To maximize the spin–charge interconversion, we choose the orientation of the d-wave ``flower''

 as shown in Figs. \ref{fig:spinsplitter_geometries} and \ref{fig:inverse_spinsplitter_geometries}. Generalization to other orientations is straightforward and will not be considered here.}

\section{The spin–splitter effect in an altermagnetic strip}
\label{sec:SSE}

\begin{figure*}[!t]
    \centering
    \includegraphics[width=1.0\textwidth]{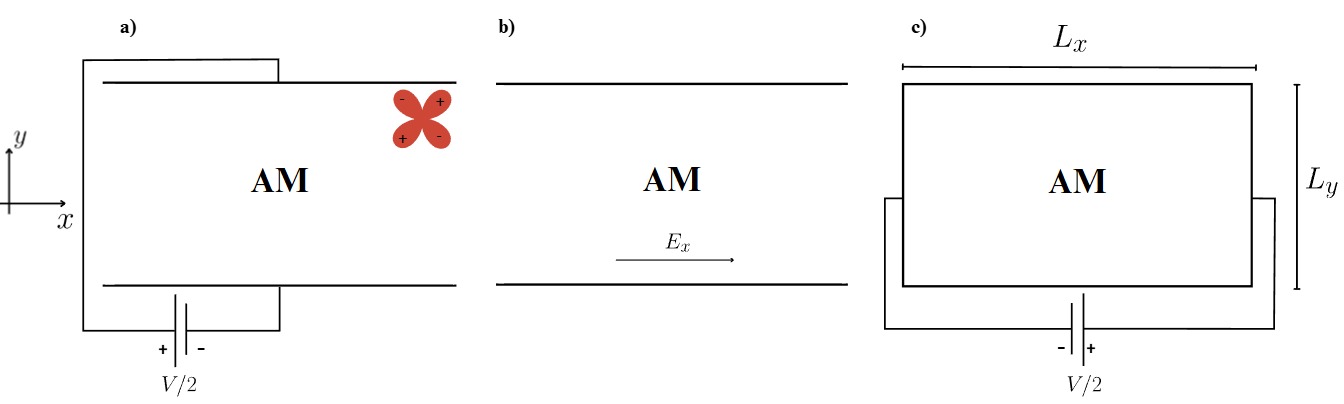}
    \caption{
    Model geometries considered for the spin–splitter effect in an altermagnetic (AM) strip. 
    (a) Transverse voltage bias configuration, where electrochemical potentials $\pm V/2$ are applied at the edges $y = \pm \frac{L_y}{2}$. 
    (b) Longitudinal electric field configuration, with a uniform field $E_x$ applied along the strip axis. 
    (c) Finite rectangular geometry of dimensions $L_x \times L_y$, combining a transverse voltage bias and open lateral boundaries. 
    These configurations serve as the basis for modeling spin–charge coupling in diffusive altermagnets.
    }
    \label{fig:spinsplitter_geometries}
\end{figure*}

In this section we analyze the spin–splitter effect in a diffusive altermagnetic  strip.  
Starting from the constitutive relations for charge and spin currents, we formulate the model describing how the spin–momentum coupling intrinsic to the altermagnet gives rise to coupled spin and charge transport under an applied electric bias. With this, we generalize the results obtained in \cite{Kokkeler2025} for a perfect contact between the altermagnet and electrode to the more realistic setting in which there is a finite barrier resistance. 
We consider three geometries, illustrated in Fig.~\ref{fig:spinsplitter_geometries}: a transverse voltage bias in an infinite strip (a), a longitudinal electric field configuration (b), and a finite rectangular sample (c).

\subsection{ Altermagnetic strip with a transverse voltage}

We first consider the configuration shown in
Fig.~\ref{fig:spinsplitter_geometries}(a): a two-dimensional altermagnetic (AM)
strip infinite along the $x$ direction and of finite width $L_y$ along $y$.
A transverse voltage $\pm V/2$ is applied at the edges
$y=\pm L_y/2$.

At the AM/normal-metal interfaces, the boundary conditions for the charge
current follow from Eq.~(\ref{eq:KLESS}) and read
\begin{equation}
\sigma_D n_y \partial_y \mu
=
G_B \left(\mu_{\mathrm{res}}-\mu\right)\;.
\end{equation}

Due to translational invariance along $x$, both the electrochemical potential
$\mu$ and the spin accumulation $\mu^s$ depend only on the transverse coordinate
$y$.
Solving for $\mu(y)$ yields
\begin{equation}
\mu(y)
=
\frac{G_B V}{G_B L_y + 2\sigma_D}\,y\;.
\end{equation}
The corresponding charge current is
\begin{equation}
j_y
=
-
\frac{G_B(\sigma_D/L_y)}
{G_B + 2(\sigma_D/L_y)}\,V\;,
\end{equation}
which reflects a series combination of the two interface conductances $G_B$ and
the transverse conductance of the altermagnet $\sigma_D/L_y$.
At this order, the spin accumulation vanishes, $\mu^s=0$.

According to Eq.~(\ref{eq:currents}), a charge current induces a spin current via
the spin-splitter effect,
\begin{equation}
j_x^s
=
T_{xy}\,j_y\;.
\end{equation}
In the limit of large interface conductance,
$G_B \gg \sigma_D/L_y$, this expression reduces to
\begin{equation}
j_x^s
=
-\sigma_D T_{xy}\,\frac{V}{L_y}\;,
\end{equation}
in agreement with Ref.~\cite{Kokkeler2025}.

\subsection{Longitudinal voltage}

We now consider a finite AM with rectangular geometry of length $L_x$
and width $L_y$, as shown in Fig.~\ref{fig:spinsplitter_geometries}(c).
The altermagnet is connected at $x=0$ and $x=L_x$ to two normal-metal reservoirs
biased at voltages $\pm V/2$.
The transverse boundaries at $y=\pm L_y/2$ are assumed to be insulating for both
charge and spin currents.

Charge and spin transport within the altermagnet are described by the
electrochemical potential $\mu(x,y)$ and the spin accumulation $\mu^s(x,y)$.
Using the drift--diffusion framework introduced previously
Eqs.~(\ref{eq:chargediffnormalAM})–(\ref{eq:currents}),
the coupled diffusion equations take the form
\begin{align}
(\partial_x^2+\partial_y^2)\mu
+2T_{xy}\partial_x\partial_y \mu^s &= 0\;,
\label{eq:mu_finite_GB}\\
(\partial_x^2+\partial_y^2)\mu^s
+2T_{xy}\partial_x\partial_y \mu &= \frac{\mu^s}{\ell_s^2}\;.
\label{eq:mus_finite_GB}
\end{align}
In the following, we treat the spin--charge coupling parameter $T_{xy}$
perturbatively and retain only terms up to linear order in $T_{xy}$.

At the interfaces with the elctrodes,  $x=0,\;L_x$, the boundary conditions follow
from  Eq.~(\ref{eq:KLESS}).
For the charge sector, they read
\begin{equation}
\sigma_D n_x \partial_x \mu = G_B (\mu_{\mathrm{res}} - \mu)\;,
\qquad
x=0,L_x\;,
\label{eq:interface_charge}
\end{equation}
whereas for the spin sector (the normal-metal reservoirs are assumed not to sustain any
spin accumulation,  $\mu^s_{\mathrm{res}}=0$)
 Eq.~(\ref{eq:KLESS}) yields
\begin{equation}
\sigma_D \, n_x \partial_x \mu^s
=
- G_B \, \mu^s\;,
\qquad
x=0,L_x\;,
\label{eq:BC_spin_longitudinal}
\end{equation}
where $n_x$ denotes the outward normal to the altermagnet.
At the transverse boundaries $y=\pm L_y/2$, the y-components of both charge and spin currents vanish.

To solve the resulting boundary-value problem, we proceed perturbatively in $T_{xy}$. At zeroth  order, the electrochemical potential follows from the diffusion equation. Eq. (\ref{eq:mu_finite_GB}),  together with the interface boundary conditions, Eq.  (\ref{eq:interface_charge}), with $\mu_{\text{res}}=\pm V/2$ results in 
\begin{equation}
\mu(x)
=
\frac{G_B}{G_B L_x + 2\sigma_D}
\left(x-\frac{L_x}{2}\right)V\;.
\label{eq:mu_GB}
\end{equation}
For highly transparent interfaces, $G_B \gg \sigma_D/L_x$, this expression reduces
to 
$\mu = V(x-L_x/2)/L_x$.

According to the previous section, the finite longitudinal gradient of $\mu$ induces a transverse spin current via
the spin-splitter effect.
 At $y=\pm L_y/2$ we imposes, vanishing spin current, which according to Eq. (\ref{eq:KLESS})  leads to following B.C:
\begin{equation}
\left.\partial_y \mu^s \right|_{y=\pm L_y/2}
=
-\,T_{xy}\,
\frac{G_B V}{G_B L_x + 2\,\sigma_D}\;.
\label{eq:BC_spin_transverse}
\end{equation}

Up to linear  order in $T_{xy}$, the spin accumulation obeys the diffusion equation Eq. (\ref{eq:mus_finite_GB})
\begin{equation}
(\partial_x^2+\partial_y^2)\mu^s
=
\frac{\mu^s}{\ell_s^2}\;,
\label{eq:mus_diffusion}
\end{equation}
subject to the boundary conditions, Eqs. ~(\ref{eq:BC_spin_longitudinal},\ref{eq:BC_spin_transverse}).
The resulting spin accumulation can be written in closed form as
\begin{widetext}
\begin{equation}
\begin{aligned}
\mu^s(x,y)
&=
-\,T_{xy}\,
\frac{V\ell_s}{L_x + 2\sigma_D/G_B}
\left(
\frac{\sinh\frac{y}{\ell_s}}{\cosh\frac{L_y}{2\ell_s}}
+\frac{4}{L_y\ell_s}\sum_{m=0}^{\infty}
\frac{1}
     {k_m^2}
\right.\\
&\qquad\left.
\frac{G_B}
{\sigma_D k_{m}\sinh\!\left(k_{m}\frac{L_x}{2}\right)
+G_B\cosh\!\left(k_{m}\frac{L_x}{2}\right)}
\cosh\!\left[k_{m}\!\left(x-\frac{L_x}{2}\right)\right]
\cos\!\left[\frac{(2m+1)\pi}{L_y}
\left(y+\frac{L_y}{2}\right)\right]
\right)\;,
\end{aligned}
\label{eq:mus_finite_correct}
\end{equation}
\end{widetext}
where
\begin{equation}
k_{m}^2
=
\ell_s^{-2}
+
\frac{(2m+1)^2\pi^2}{L_y^2}\;.
\label{eq:kvector}
\end{equation}

\begin{figure}[h]
    \centering
    \includegraphics[width=1.05\linewidth]{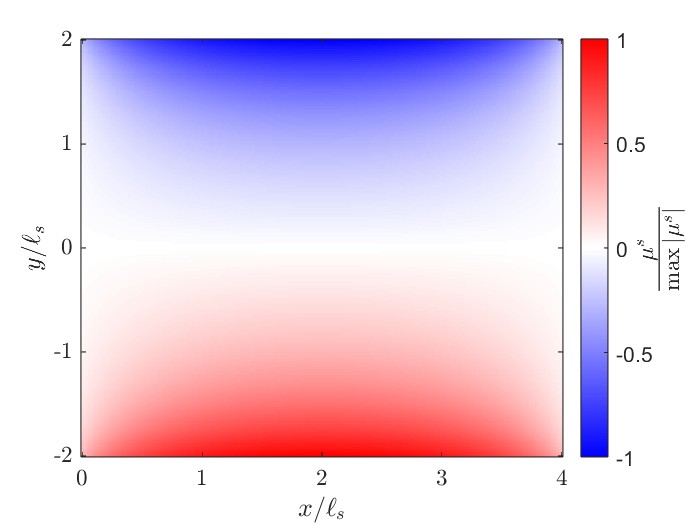}
    \caption{Spin accumulation in a finite altermagnetic strip induced by a longitudinal voltage difference, see Fig.~\ref{fig:spinsplitter_geometries}(c).
    The dimensions of the strip are $L_x/\ell_s = L_y/\ell_s = 4$.
    The plot is obtained from Eq.~(\ref{eq:mus_finite_correct}) for $ G_B\ell_s/\sigma_D = 1.5$, illustrating the spatial profile of the spin accumulation. 
    }
    \label{fig:placeholder}
\end{figure}

Equation~(\ref{eq:mus_finite_correct}) corresponds to a Fourier mode expansion that
automatically satisfies the transverse boundary conditions. The corresponding spatial spin profile   is shown in Fig.~\ref{fig:placeholder}, illustrating the distribution of the spin in the finite AM.

One can verify that in the limit of perfectly transparent interfaces, $G_B \to \infty$,  Eq. (\ref{eq:mus_finite_correct}) coincides with the result presented in Ref. \cite{Kokkeler2025}. 

Another limiting case is the infinite strip in the $x$ direction, 
Fig.~\ref{fig:spinsplitter_geometries}(b).
Because of translational invariance along $x$, the spin accumulation depends
only on $y$, and it is given by the first term in parentheses in Eq. (\ref{eq:mus_finite_correct}):
\begin{equation}
\mu^s(y)
=
-\,T_{xy}\ell_s\,E_x\,\,
\frac{\sinh(y/\ell_s)}
{\cosh(L_y/2\ell_s)}\;,
\label{eq:mus_infinite}
\end{equation}
where $E_x$
 is the homogeneous electric field  along the strip, see Fig.~\ref{fig:spinsplitter_geometries}(b).
The spin accumulation at the edges follows directly from
Eq.~(\ref{eq:mus_infinite}) and is given by
\begin{equation}
\mu^s\!\left(\pm\frac{L_y}{2}\right)
=
\pm\,T_{xy}\ell_s\,E_x\,
\tanh\!\left(\frac{L_y}{2\ell_s}\right)\;.
\label{eq:edge_spin}
\end{equation}
This spin accumulation can be detected directly using magnetic electrodes and
non-local spin-valve measurements, as discussed in
Sec.~\ref{sec:hybrid}, or indirectly via SMR-type measurements
\cite{leiviska2025spin,chen2025spin}.

\section{The inverse spin-splitter effect}
\label{sec:ISSE}

\begin{figure*}[t!]
    \centering \includegraphics[width=1.03\textwidth]{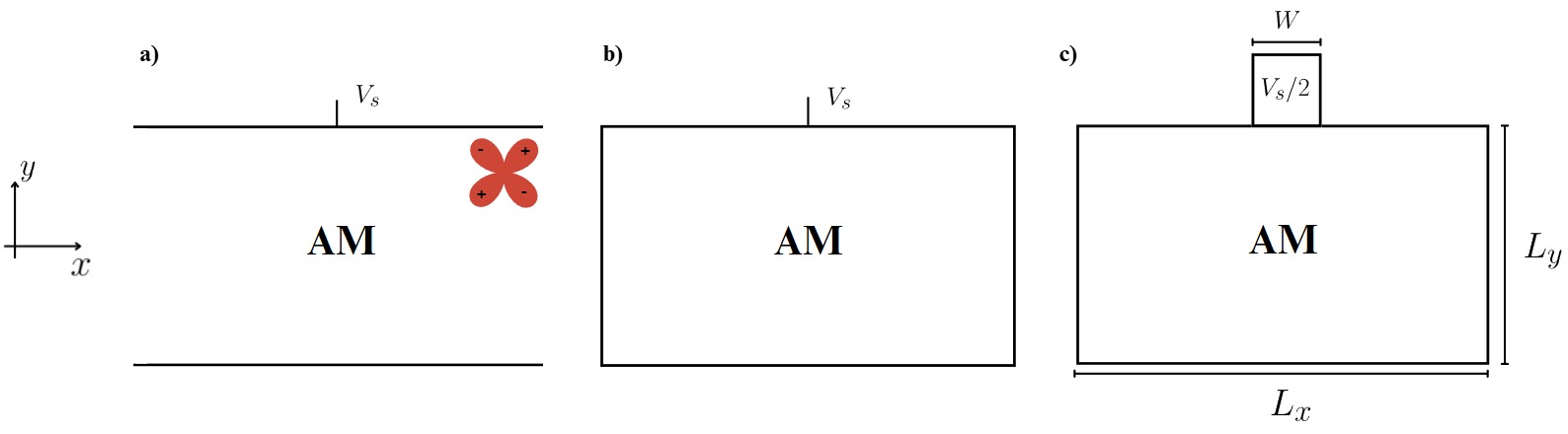}
    \caption{
   Geometries considered for transverse spin injection in an
    altermagnetic (AM) strip.
    (a) Uniform spin injection at the upper transverse boundary of an infinite
    strip, assuming translational invariance along the longitudinal direction.
    (b) Spin injection at the upper transverse edge of a finite strip, with
    transverse boundaries located at $y=\pm L_y/2$.
    (c) Spin injection localized over a finite region of the
    upper transverse edge of a finite strip.
    }
    \label{fig:inverse_spinsplitter_geometries}
\end{figure*}

Following Eq. (\ref{eq:currents}), the  tensor $T_{jk}$ enters the expressions for the charge current and spin current analogously. This suggests that the spin-splitter effect discussed in the previous section, in which a voltage generates a spin accumulation, should be accompanied by a reciprocal response, the generation of a charge accumulation via a spin voltage, which we term the inverse spin-splitter effect.
In this section we analyze the inverse spin--splitter effect in a diffusive
altermagnetic strip.
We consider three  spin-injection geometries, illustrated in
Fig.~\ref{fig:inverse_spinsplitter_geometries}: a uniform spin accumulation
applied to an infinite strip (a), spin injection into a finite strip  (b), and a local spin injection
in  a finite sample (c).

\subsection{Infinite altermagnetic strip}

We first  consider an altermagnetic strip that is infinite along the $x$--direction and
has a finite width $L_y$ along $y$, as illustrated in
Fig.~\ref{fig:inverse_spinsplitter_geometries}(a).
This configuration is reciprocal to the one studied in
Sec.~\ref{sec:SSE}: instead of applying an electric bias, a transverse
\emph{spin accumulation} (or spin voltage) is imposed at {$y=L_y/2$}, while charge transport across the boundaries is suppressed.
Such a situation can be realized, for instance,  by spin pumping, or in nonlocal spin--valve geometries
where a pure spin current is injected into the altermagnet.

Because the system is translationally invariant in the $x$-direction, $\mu^s$ depends only on $y$, satisfying, in the stationary regime, the spin diffusion equation  
\begin{equation}
\partial_y^2 \mu^s = \frac{\mu^s}{\ell_s^2}\;.
\label{eq:spin_diffusion_ISS}
\end{equation}
At the upper edge, see Fig.~\ref{fig:inverse_spinsplitter_geometries}(a),  $y= L_y/2$, the boundary condition,  Eq.~(\ref{eq:KLESS}), reads
\begin{equation}
\sigma_D n_y \partial_y \mu^s
=
G_B\bigl(V_s-\mu^s\bigr)|_{y=\frac{L_y}{2}}\;,
\label{eq:BC_spin_ISS}
\end{equation}
whereas at $y=-L_y/2$ we impose a zero spin current, {\it i.e.} $\partial_y \mu^s=0$. 

The solution of  Eq.~(\ref{eq:spin_diffusion_ISS}) together with these boundary conditions is given by
\begin{equation}
\mu_s(y)
=\,
\frac{V_s}
{ 1+\dfrac{\sigma_D}{G_B\ell_s}
\tanh\!\left(\dfrac{L_y}{\ell_s}\right) }
\,
\frac{\cosh\!\left(\dfrac{y+L_y/2}{\ell_s}\right)}
{\cosh\!\left(\dfrac{L_y}{\ell_s}\right)} \; .
\label{eq:mus_ISS}
\end{equation}

According to   Eq.~(\ref{eq:currents}), the  spin current, proportional to the spin gradient  leads  to a longitudinal  charge current via the inverse spin-splitter effect:
\begin{equation}
j_x = -\sigma_D T_{xy}\partial_{y}\mu^s\;.
\label{eq:jx_ISS}
\end{equation}
Substitution of  Eq.~(\ref{eq:mus_ISS}), leads to
\begin{equation}
j_x(y)=
-\frac{\sigma_D T_{xy}V_s}{\ell_s} 
\frac{\sinh\left(\dfrac{y+L_y/2}{\ell_s}\right)}
{\left[ 1+\dfrac{\sigma_D}{G_B\ell_s}
\tanh\left(\dfrac{L_y}{\ell_s}\right)\right] \cosh\left(\dfrac{L_y}{\ell_s}\right)}
\,.
\label{eq:jx_profile_ISS}
\end{equation}
The total longitudinal charge current
induced by the imposed spin accumulation is obtained by integration over $y$: 
\begin{equation}
I_{\text{ISS}}
=-
\frac{\sigma_DT_{xy}V_s}{ 1+\dfrac{\sigma_D}{G_B\ell_s}\tanh\left(\frac{L_y}{\ell_s}\right)} \frac{\cosh\left(\frac{L_y}{\ell_s}\right) -1}{\cosh\left(\frac{L_y}{\ell_s}\right)}\;.
\label{eq:Ix_ISS}
\end{equation}
In the limit of large interface conductance,
$G_B \gg \sigma_D/\ell_s$, Eq.~(\ref{eq:Ix_ISS}) simplifies to
\begin{equation}
I_{\text{ISS}} = -G_B \ell_sT_{xy}\frac{\cosh\left(\frac{L_y}{\ell_s}\right) -1}{\sinh\left(\frac{L_y}{\ell_s}\right)}
\label{eq:Ix_ISS_largeGB}
\end{equation}


Let us now assume that the stripe is also finite in $x$ direction, as shown in Fig. \ref{fig:inverse_spinsplitter_geometries}(b). In this case the
total average current is zero,  $I_x=\int dy j_x=0$.  Hence, from Eq. (\ref{eq:currents}) we obtain:
\begin{equation}
   \frac{1}{L_y}\int dy \partial_x \mu=-\frac{T_{xy}}{L_y} \Delta \mu^s\; , 
\end{equation}
where $\Delta \mu^s$ is the difference of the spin in the upper and lower edges.  
If we now integrate this equation with respect to $x$ we obtained the average  voltage between the transverse edges. Up to  linear order in $T_{xy}$,  we can use the solution Eq.~(\ref{eq:mus_ISS}) for $\mu_s$ resulting in: 
\begin{equation}
    V_{\text{ISS}}=-\frac{T_{xy}L_x}{L_y}\frac{V_s}
{ 1+\dfrac{\sigma_D}{G_B\ell_s}
\tanh\left(\dfrac{L_y}{\ell_s}\right) }\frac{\cosh\frac{L_y}{\ell_s}-1}{\cosh\frac{L_y}{\ell_s}}\;.
\end{equation}
This is the transverse voltage generated by spin injection in an open geometry.  As expected,  corresponds to 
the expression for the transverse current $I_{\text{ISS}}$, Eq. (\ref{eq:Ix_ISS}), divided by the effective transverse conductance, $\sigma_D L_y/L_x$.

\subsection{Local spin injection in an altermagnetic strip} 
\label{sec:ISS_point}


\begin{figure}[h]
    \centering
    \includegraphics[width=1.0\linewidth]{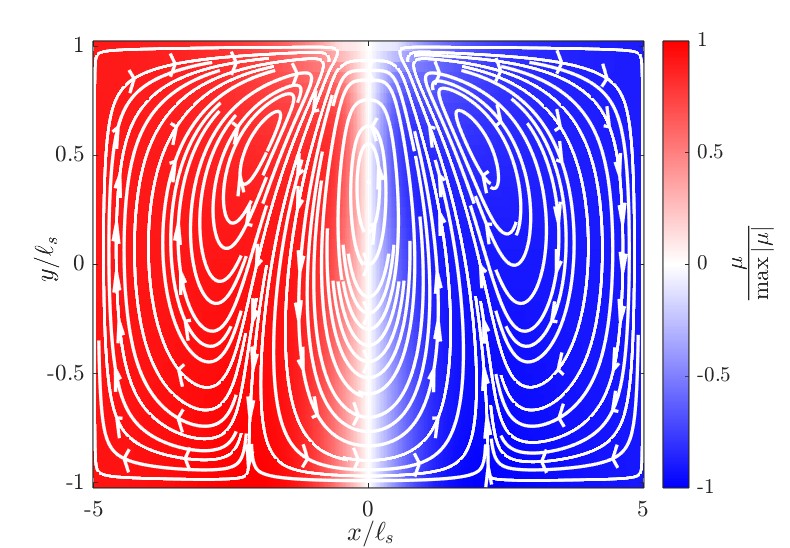}
    \caption{Normalized electrochemical potential $\mu(x,y)/\max|\mu|$ together with the induced charge
current streamlines generated by localized Gaussian spin injection at the upper edge of a
finite altermagnetic strip.
Spatial coordinates are given in units of the spin relaxation length $\ell_s$.
The parameters used are $L_x/\ell_s=10$, $L_y/\ell_s=2$ and a Gaussian
injection profile of width $W=0.25\,\ell_s$ centered at $x=0$.}
    \label{fig:spininjection}
\end{figure}

\begin{figure}[t!]
    \centering
    \includegraphics[width=1\linewidth]{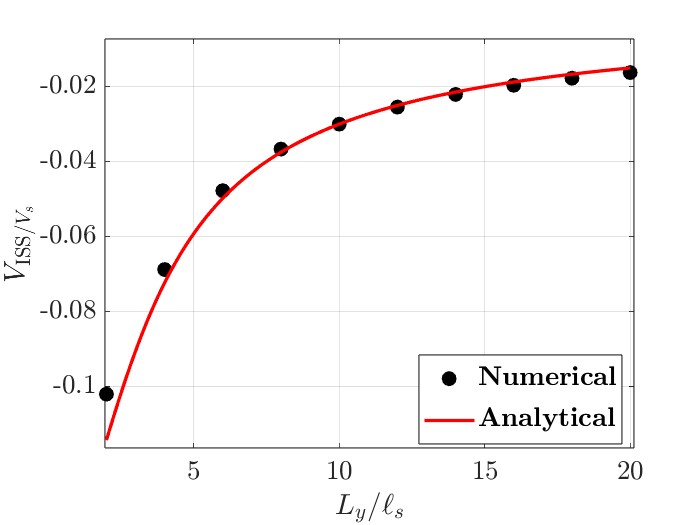}
    \caption{Induced inverse spin--splitter voltage $V_{\mathrm{ISS}}$ as a function of the transverse strip
length $L_y$.
Symbols correspond to the numerical results obtained from the full diffusive calculation,
while the solid line shows the semi--analytical prediction given by Eq.~(\ref{eq:VISS_tanh}).
Parameters used are $G_B W/\sigma=0.15$, $T_{xy}=1$, and $L_x/\ell_s=40$. }
    \label{fig:vissspininjection}
\end{figure}

Thus far we have only considered geometries in which the contact between the the injector and the altermagnet is wide, extending along the whole edge. Another geometry that can be realized is that of a local injector. To study this type of system, we consider the configuration illustrated in
Fig.~\ref{fig:inverse_spinsplitter_geometries}(c).

Charge and spin transport in the altermagnet are governed by the same
drift--diffusion equations introduced in the previous sections
Eqs.~(\ref{eq:mu_finite_GB}) and (\ref{eq:mus_finite_GB}), together with the
constitutive relations of the model Eqs. (\ref{eq:currents}) and (\ref{eq:KLESS}).
In the present geometry, the essential new ingredient is the localized
character of the spin injection, which enters through the boundary
conditions at the upper transverse edge of the strip.

We consider spin injection through an injector of width $W$ satisfying
$W\ll \ell_s, L_x, L_y$. In this regime, the injector can be approximated as
point-like and is described by a Dirac $\delta$-function in the boundary
conditions.

The boundary conditions along the transverse direction read
\begin{align}
&\partial_y \mu_s^z+ T_{xy}\partial_x\mu\big|_{x,y=-L_y/2}
= 0\;,
\label{eq:inj_bc_bottom}
\\
\begin{split}
    &\sigma_D\!\left(
\partial_y \mu_s^z
+ T_{xy}\partial_x\mu
\right)|_{x,y = L_y/2}
= \\
&G_B\,W\,\delta(x)\bigl(V_s-\mu_s^z(x,\frac{L_y}{2})\bigr)\;.
\end{split}
\label{eq:inj_bc_top}
\end{align}
Here, $W$ is the effective width of the injector and $V_s$ is the applied
spin voltage. The Dirac delta explicitly encodes the localized nature of
the injection, while the term proportional to $\mu_s^z$ accounts for the
back-action of the induced spin accumulation at the interface.

At $x = \pm \frac{L_x}{2}$, we impose vanishing currents normal to the interface,
\begin{equation}
    \partial_x \mu_s^z(\pm \frac{L_x}{2},y)
    + T_{xy}\partial_y\mu(\pm \frac{L_x}{2},y)
    = 0\;.
    \label{eq:inj_bc_sides}
\end{equation}

Although a closed analytical solution for the full spatial profiles
$\mu(x,y)$ and $\mu_s^z(x,y)$ is not available, an analytical expression
for the ISS-induced voltage along the $x$-direction can be obtained by focusing on the spin
accumulation averaged over the entire sample area. As shown in
Appendix~\ref{app:VISS2}, the induced voltage can be written as
\begin{equation}
    V_{\mathrm{ISS}}
    =
    -T_{xy}\,\frac{L_x}{\ell_s}\,
    \langle \mu_s \rangle\,
    \tanh\!\left(\frac{L_y}{2\ell_s}\right)\;,
    \label{eq:VISS_tanh}
\end{equation}
where $\langle \mu_s \rangle$ denotes the spin potential averaged over the
full two-dimensional geometry of the altermagnetic strip.

To evaluate $\langle \mu_s \rangle$, we integrate the diffusion equation
for $\mu_s^z$ over the entire spatial domain and make use of the boundary
conditions Eqs.~(\ref{eq:inj_bc_bottom}) and (\ref{eq:inj_bc_top}). This yields,
to lowest order in $T_{xy}$ (see Appendix~\ref{app:SASA} for details),
\begin{equation}
    \langle \mu_s \rangle
    =
    \frac{G_B \, W\ell_s^2}{\sigma_DL_x L_y}
    \bigl(V_s-\mu_s^z(0,\frac{L_y}{2})\bigr)\;.
    \label{eq:mu_avg_exact}
\end{equation}


Let us consider the case in which the interface conductance is small $G_B W\ll \sigma$. Since $\mu_s$ vanishes in the absence of the electrode, the term
$G_B W \mu_s^z(0,\frac{L_y}{2})$ contributes only at order $\mathcal{O}((G_B W/\sigma)^2)$ in Eq. (\ref{eq:mu_avg_exact}) 
and can therefore be neglected at leading order. With this, Eq.~\eqref{eq:mu_avg_exact} reduces to (see Appendix~\ref{app:SASA} for details)
\begin{equation}
    \langle \mu_s \rangle
    \approx
    \frac{2G_B \, W\ell_s^2}{\sigma_D L_x L_y}\,V_s\;.
    \label{eq:mu_avg_first_order}
\end{equation}
Together with Eq.~\eqref{eq:VISS_tanh}, this provides a closed-form
expression for $V_{\text{ISS}}$ in the weak-coupling regime.

To validate the analytical result and characterize the full spatial
response, Eqs.~(\ref{eq:mu_finite_GB}) and (\ref{eq:mus_finite_GB})  are solved numerically together with the boundary
conditions Eqs. ~(\ref{eq:inj_bc_bottom}) and (\ref{eq:inj_bc_top}). The numerical
solutions yield the stationary profiles of $\mu(x,y)$ and
$\mu_s^z(x,y)$, from which the induced charge current density is
reconstructed.

Figure~\ref{fig:spininjection} shows the electrochemical potential
$\mu(x,y)$ together with the corresponding charge current streamlines.
Three distinct current loops emerge as a consequence of the anisotropic
transport tensor of the altermagnet combined with the localized nature
of the spin injection, providing a clear manifestation of the inverse
spin--splitter mechanism.  Such current loops have also been obtained in Ref.~\cite{Linder2} in other setups, for which the authors claim they can be measured using magnetometry techniques. Here, in contrast, we focus on electric measurements of the transverse electrical signals in conventional transport experiments using multiterminal devices.

The inverse spin--splitter voltage is obtained from the numerical
solution as the difference between the electrochemical potentials
averaged over the transverse direction at the two lateral edges of the
strip. The resulting $V_{\mathrm{ISS}}$ is shown in
Fig.~\ref{fig:vissspininjection} as a function of the transverse length
$L_y$. Symbols correspond to the numerical results, while the solid line
represents the analytical prediction derived within the first-order
expansion in the barrier conductance $G_B$ in Eqs. (\ref{eq:VISS_tanh}) and (\ref{eq:mu_avg_first_order}). The agreement demonstrates that the leading-order analytical treatment
accurately captures the essential features of the inverse
spin--splitter effect induced by point-like spin injection.

\bigskip

In short, the results presented in this section establish the inverse
spin--splitter effect as the reciprocal counterpart of the spin--splitter
effect discussed in Sec.~\ref{sec:SSE}. In both cases, the conversion mechanism is
governed by the same nonrelativistic spin--momentum coupling intrinsic to the
altermagnetic state and does not rely on (relativistic) spin--orbit interaction. The
analytical and numerical results obtained for extended and localized spin
injection geometries demonstrate that the inverse response is robust against
details of the injection profile and interface transparency, providing a
consistent and experimentally accessible signature of altermagnetic
spin--charge conversion in hybrid nanostructures.

\section{Hybrid multi-terminal setups}
\label{sec:hybrid}

In the previous sections, we analyzed how charge--spin interconversion arises
in altermagnets within a generalized drift--diffusion framework. In this
section, motivated by the recent non-local spin-valve experiment reported in
Ref.~\cite{mencos2025direct}, we study how this conversion manifests itself in
hybrid multi-terminal transport geometries. In particular, we show that a
charge current flowing along an altermagnetic strip generates a non-local
electrical signal in an attached normal metal through spin injection mediated
by the spin--splitter effect. The magnitude and sign of the resulting non-local
voltage are controlled by the orientation of the Néel vector, establishing a
direct connection between altermagnetic order and non-local spin transport.

\begin{figure}[h]
    \centering
    \includegraphics[width=1.0\linewidth]{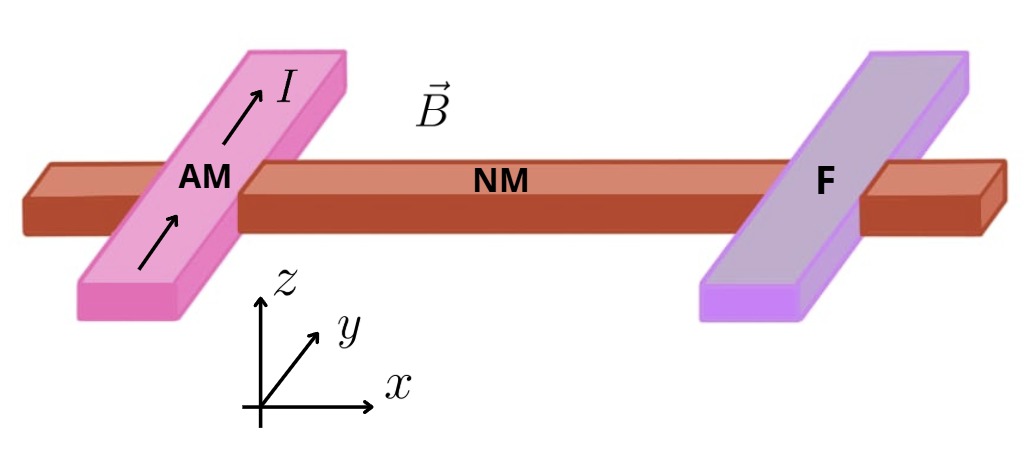}
    \caption{
Hybrid nonlocal spin valve. A finite altermagnetic strip injects spin into a
diffusive normal metal, which is electrically detected by a ferromagnetic
(F) electrode. The induced spin can be rotated by  an external magnetic field $\vec B$.
}
\label{fig:sistemahibrido}
\end{figure}

We consider a typical  non-local spin valve \cite{jedema2001electrical,villamor2015modulation,ji2007non}  shown in Fig.~\ref{fig:sistemahibrido},
consisting of a diffusive normal metal (NM) attached to an altermagnetic 
strip and a ferromagnetic detector. In the configuration considered here, the
driving charge current flows exclusively along the altermagnetic arm, while the
normal metal and ferromagnetic electrodes act solely as spin transport and
detection elements. When a charge current flows along the AM arm, a transverse
spin current is generated via the spin--splitter effect discussed in
Section~\ref{sec:SSE}. This spin current is injected into the NM and diffuses
over the spin diffusion length, which in metals such as Cu can reach micrometer
scales. The resulting spin accumulation is detected electrically by a
ferromagnetic electrode placed at a distance from the injector. The geometry 
corresponds to the one of the electrical bias geometry of a recent experiment reported in
Ref.~\cite{mencos2025direct}. In what follows, we compute the resulting non-local
voltage signal. In addition to the zero-field configuration, we also consider an
external magnetic field applied along the NM wire, which induces spin
precession and gives rise to a Hanle-type response.

In the stationary diffusive regime, spin transport in the NM is governed by
\begin{equation}
\partial_x^2 \vec{\mu}_s
-
\frac{\vec{\mu}_s}{\ell_{NM}^2}
+
\chi\,(\vec{B}\times\vec{\mu}_s)
=0\;,
\end{equation}
where $\vec{\mu}_s=(\mu_s^x,\mu_s^y,\mu_s^z)$ denotes the spin accumulation and
$\ell_{NM}$ is the spin diffusion length. The parameter
$\chi=g\mu_B/D$ depends on the effective gyromagnetic factor $g$, the Bohr
magneton $\mu_B$, and the diffusion constant $D$ of the normal metal.

The spin injected from the altermagnet is polarized along the Néel vector
$\vec N$ of the AM,
\begin{equation}
\vec{\mu}_s^{\mathrm{AM}} = \mu_s^{\mathrm{AM}}\,\vec N \;.
\end{equation}
Here $\mu_s^{\mathrm{AM}}$ is the spin accumulation at the edges of the AM strip induced by the charge current in the AM, and given by Eq. (\ref{eq:edge_spin}).
Spin injection across the AM/NM interface is described by the boundary
condition
\begin{equation}
-\vec {\hat n}\cdot\vec j_s
=
G_B\left(\mu_s^{\mathrm{AM}}-\mu_s\right)\Big|_{x=0}\;,
\end{equation}
where $G_B$ denotes the interface spin conductance.


Solving the diffusion equation with the boundary condition above yields the
spin accumulation in the NM,
\begin{equation}
\begin{split}
\vec{\mu}_s(x)
&=
\frac{G_B}{G_B+\sigma_N\alpha}\,
\mu_s^{\mathrm{AM}}\,
\Big[
(\vec N\cdot\hat{\vec B})\hat{\vec B}\,e^{-x/\ell_{NM}}
\\
&\quad
+ \vec N\times\hat{\vec B}\,e^{-\alpha x}\sin(\beta x)\\
&+ \hat{\vec B} \times(\vec N\times\hat{\vec B})\,e^{-\alpha x}\cos(\beta x)
\Big]\;,
\end{split}\label{eq:Hanle_spin}
\end{equation}
where $\hat{\vec B}=\vec B/|\vec B|$, $\sigma_N$ is the conductivity of the
normal metal, and $\alpha$ ($\beta$) is the real (imaginary) part of $\sqrt{1/\ell_{NM}^2+i\chi B}$. 

A ferromagnetic detector placed at position $x=L$ measures the  non-local voltage \cite{jedema2002electrical,heikkila2019thermal}: 
\begin{equation}
V_{NL}(\vec m)=V_0+{P}\,\vec{\mu}_s(L)\cdot\vec m\;,
\label{eq:VNL}
\end{equation}
where $V_0$ is a spin-independent voltage, and  $P$ is the  polarization  of the F detector whose polarization is parallel to  the unit vector  $\vec m$.

 The relevant observable is the voltage difference, $\Delta V_{NL}
\equiv V(\hat{\vec m})-V(-\hat{\vec m})$, measured for
opposite orientations of the detector magnetization.  From Eqs. (\ref{eq:Hanle_spin}) and (\ref{eq:VNL}) we obtain:
\begin{equation}
\begin{split}
\Delta V_{NL}=
\frac{2P\,G_B \mu_s^{\mathrm{AM}}}{(G_B+\sigma_N\alpha)}&\,
\,
\hat{\vec m}\cdot
\Big[
(\vec N\cdot\hat{\vec B})\hat{\vec B}\,e^{-L/\ell_{NM}}\\
 +\vec N\times\hat{\vec B}&\,e^{-\alpha L}\sin(\beta L)\\
+\hat{\vec B}\times&(\vec N \times\hat{\vec B})\,e^{-\alpha L}\cos(\beta L)
\Big]\;.
\end{split}
\end{equation}
Let us first analyze the case of a zero Hanle  field. After substitution of Eq. (\ref{eq:edge_spin}) for $\mu_s^{\mathrm{AM}}$ we obtain:  
\begin{align}
\Delta V_{NL}=
\frac{2P\,G_B}{(G_B+\sigma_N\alpha)}\,
E_yT_{xy}\ell_s\,
\hat{\vec m}\cdot
 \vec{N}e^{-L/\ell_{NM}}\; . 
\end{align} 
Thus, the non-local voltage is proportional to the altermagnetic tensor $T_{xy}$ and maximized when the Neél vector of the AM is collinear to the magnetization of the FM.

\begin{figure}[h]
    \centering
    \includegraphics[width=1.1\linewidth]{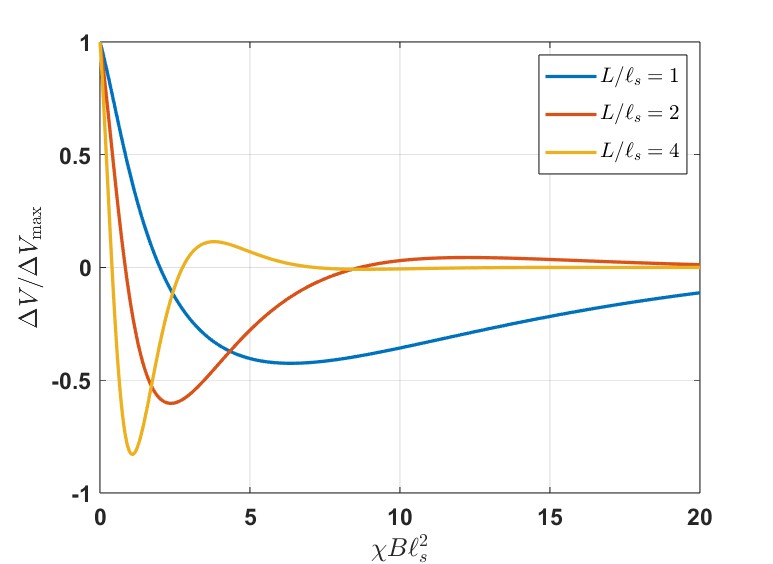}
\caption{
Normalized nonlocal voltage $\Delta V/\Delta V_{\max}$ as a function of
$\chi B \ell_{NM}^2$ for different injector--detector separations $L/\ell_{NM}=1,\,2,\,4$.
The curves are obtained from the general expression for the nonlocal signal,
evaluated for a magnetic field applied along the $z$ direction and a detector
magnetization oriented along the $y$ axis. The Néel vector is chosen in the
plane perpendicular to the field, with components $N_x=\cos\theta$ and
$N_y=\sin\theta$, using $\theta=\pi/4$. The simulation is performed in
dimensionless units with $\ell_s =\ell_{NM} =1$, so that the magnetic field
dependence enters exclusively through the combination $\chi B \ell_{NM}^2$. 
}

\label{fig:hanle}
\end{figure}

For a finite field, we consider as an illustrative example  an  external magnetic field  applied along the
$z$ axis, $\vec B = B\hat z$, while the magnetization of the ferromagnetic
detector is oriented along the transverse $y$ direction,
$\vec m = \hat y$. 
In this case we obtain 
\begin{equation}
\begin{split}
 \Delta V_{NL}=
\frac{2P\,G_B \mu_s^{\mathrm{AM}}}{(G_B+\sigma_N\alpha)}\,
\,e^{-\alpha L}\Big[
&N_x\,\sin(\beta L)\\
&+
N_y\,\cos(\beta L)
\Big]\;.   
\label{eq:nonlocalV}
\end{split}
\end{equation}

Figure~\ref{fig:hanle} shows the corresponding non local Hanle curves obtained
from Eq.~(\ref{eq:nonlocalV}) for this particular geometry, highlighting the precession and
dephasing of the spin accumulation generated by the altermagnet as the magnetic
field is varied, for different injector--detector separations.

\section{Conclusions}\label{sec:conclusions}
Building on a kinetic-equation approach, we formulate coupled drift–diffusion equations to describe charge and spin transport in hybrid devices containing altermagnetic materials.

We first analyze the spin-splitter effect, namely the conversion of charge currents into spin currents, in diffusive altermagnetic strips for several device geometries. We then investigate the reciprocal phenomenon—the inverse spin-splitter effect—where an injected spin accumulation generates a transverse electrical response in the altermagnet. For both extended and localized spin injection, we derive closed-form expressions for the induced voltage in the weak-coupling regime and validate them through numerical solutions of the coupled diffusion equations.

Motivated by a recent experiment, we also apply our theoretical framework to hybrid multiterminal geometries in which an altermagnetic strip acts as a spin injector for a diffusive normal metal. We demonstrate that the spin accumulation generated by the spin-splitter effect in the altermagnet and injected in the normal wire,  can be efficiently detected using a conventional ferromagnetic electrode in a nonlocal configuration. The measured nonlocal voltage is proportional to the scalar product of the Néel vector of the altermagnet and the polarization of the ferromagnetic detector.

Finally, we analyze the effect of an external magnetic field in the same geometry. The resulting Hanle-type signal exhibits the characteristic oscillatory behavior associated with spin precession during diffusive transport, providing another  direct electrical probe of spin injection from altermagnets.

Our theory establishes altermagnets as robust and electrically controllable sources of spin in multiterminal spintronic devices and can be straightforwardly adapted to a wide variety of hybrid structures combining altermagnets, normal metals, and ferromagnets.

\section*{Acknowledgements}

We thank I. V. Tokatly, F. Casanova and M. Borra for useful discussions and support in this work.
N.S. and G.D.P. acknowledges support from the Programa de Desarrollo de las Ciencias Básicas (PEDECIBA) and from the grant of number FCE-3-2024-1-180709 of the Agencia Nacional de Investigación e Innovación (Uruguay).
The work of T.K.  was supported by the Research Council of Finland through DYNCOR, Project Number 354735 and through the Finnish Quantum Flagship, Project Number 359240. His work is part of the Finnish Centre of Excellence in Quantum Materials (QMAT). 
F.S.B  acknowledges financial support from the Spanish MCIN/AEI/10.13039/501100011033 
through grant
PID2023-148225NB-C31, and from the European Union’s Horizon Europe 
through grant JOSEPHINE (No. 101130224).

\appendix
\section{Derivation of the inverse spin--splitter voltage for point-like spin injection}
\label{app:VISS}

In this Appendix we outline the derivation of Eq. (\ref{eq:mu_avg_exact}), for the average spin accumulation in the setup, its expansion to first order in $G_BW/\sigma$, Eq. (\ref{eq:mu_avg_first_order}), and  Eq. (\ref{eq:VISS_tanh}) for the inverse spin--splitter
voltage $V_{\mathrm{ISS}}$ induced by a localized spin injection at the
boundary of a finite altermagnetic strip.

We consider a two-dimensional altermagnet occupying the region
$x\in[-\frac{L_x}{2},\frac{L_x}{2}]$ and $y\in[-\frac{L_y}{2},\frac{L_y}{2}]$.
To leading order in the spin--momentum coupling $T_{xy}$, charge and spin
potentials obey the kinetic diffusion equations, Eqs. (\ref{eq:mu_finite_GB},\ref{eq:mus_finite_GB}) in the main text, which read
\begin{align}
(\partial_x^2+\partial_y^2)\mu + 2T_{xy}\partial_x\partial_y\mu^s &= 0\;,
\label{eq:app_mu}\\
(\partial_x^2+\partial_y^2)\mu^s + 2T_{xy}\partial_x\partial_y \mu &= \frac{\mu^s}{\ell_s^2}\;,
\label{eq:app_mus}
\end{align}
%
where $\ell_s=\sqrt{D\tau_s}$ is the spin relaxation length.

Spin is injected locally at the upper boundary $y=\frac{L_y}{2}$ through a narrow
contact of width $W\ll \ell_s,L_{x,y}$, which we approximate using a $\delta$-function.
The boundary conditions Eqs. (\ref{eq:inj_bc_top},\ref{eq:inj_bc_bottom},\ref{eq:inj_bc_sides}) for the spin sector read
\begin{align}
\partial_y \mu^s(x,-\frac{L_y}{2}) + T_{xy}\partial_x \mu &= 0\;,
\label{eq:app_bc_bottom}\\
\sigma_D(\partial_y \mu^s(x,\frac{L_y}{2})+T_{xy}\partial_x \mu)
&=
2G_BW\,\delta(x)\,
\bigl[V_s-\mu^s(x,\frac{L_y}{2})\bigr]\;,\label{eq:app_bc_top}\\
\partial_x \mu^s(\pm \frac{L_x}{2},y) + T_{xy}\partial_y \mu &= 0\;.
\label{eq:app_bc_leftright}
\end{align}
Charge currents vanish at all boundaries, that is,
\begin{align}
    \partial_x \mu (\pm \frac{L_x}{2},y) = 0\;,\label{eq:app_Nochargecurrentsx}\\
    \partial_y \mu (x,\pm \frac{L_y}{2}) = 0\;.\label{eq:app_Nochargecurrentsy}
\end{align}

\subsection{Spatially averaged spin accumulation}
\label{app:SASA}
We treat the system perturbatively in $T_{xy}$. First we consider only $\mu^s$ and derive Eq. (\ref{eq:mu_avg_exact}). To first order in $\mu^s$, Eqs. (\ref{eq:app_mus}-\ref{eq:app_bc_leftright}) read
\begin{align}
    (\partial_{xx}+\partial_{yy})\mu^s &= \frac{1}{\ell_s^2}\mu^s\;\label{eq:app_mus0}\;,\\
    \partial_y \mu^s(x,-\frac{L_y}{2})&= 0\;,
\label{eq:app_bc_bottom0}\\
\sigma_D(\partial_y \mu^s(x,\frac{L_y}{2}))
&=
2G_B\,W\delta(x)\,
\bigl[V_s-\mu^s(x,\frac{L_y}{2})\bigr]\;,\label{eq:app_bc_top0}\\
\partial_x \mu^s(\pm \frac{L_x}{2},y) &= 0 \;.\label{eq:app_bc_leftright0}
\end{align}

The spatially averaged spin accumulation is defined as
\begin{equation}
\langle \mu^s \rangle
=
\frac{1}{L_x L_y}
\int_{-\frac{L_x}{2}}^{\frac{L_x}{2}}\!dx
\int_{-\frac{L_y}{2}}^{\frac{L_y}{2}}\!dy\;
\mu^s(x,y)\;.
\end{equation}
Exploiting Eqs. (\ref{eq:app_mus0}--\ref{eq:app_bc_leftright0}) and applying Gauss' theorem, we find
%
\begin{align}
\langle \mu^s \rangle &= \frac{\ell_s^2}{L_xL_y}
\int_{-\frac{L_x}{2}}^{\frac{L_x}{2}}\!dx
\int_{-\frac{L_y}{2}}^{\frac{L_y}{2}}\!dy\;
(\partial_{xx}+\partial_{yy})\mu^s(x,y)\nonumber\\& =\frac{\ell_s^2}{L_xL_y} \int_{\text{edges}} dl n_k\partial_k \mu^s \nonumber\\&= \frac{\ell_s^2}{L_xL_y} \int_{-\frac{L_x}{2}}^{\frac{L_x}{2}}\!dx 2\frac{G_{B}}{\sigma_D}W\delta(x)[V_s-\mu^s(x,\frac{L_y}{2})]\nonumber\\&= \frac{2G_BW\,\ell_s^2}{\sigma_D L_x L_y}
\bigl[V_s-\mu^s(0,\frac{L_y}{2})\bigr]\;.
\label{eq:app_mu_avg}
\end{align}
This corresponds to Eq. (\ref{eq:mu_avg_exact}) in the main text.

Next, we consider its weak-coupling limit and derive Eq. (\ref{eq:mu_avg_first_order}) in the main text.
In the weak-coupling limit $G_B\ll \sigma_D/\ell_s$, the spin accumulation at the
injection point $\mu^s(0,\frac{L_y}{2})$ contributes only at order
$\mathcal{O}\Big(\left(\frac{G_B \ell_s}{\sigma}\right)^2\Big)$ and can therefore be neglected at leading order.
This yields
\begin{equation}
\langle \mu^s \rangle
\simeq
\frac{2G_BW\,\ell_s^2}{\sigma_D L_x L_y}\,V_s\;.
\label{eq:app_mu_avg_final}
\end{equation}
This corresponds to Eq. (\ref{eq:mu_avg_first_order}) in the main text.

\subsection{Inverse spin--splitter voltage}
\label{app:VISS2}

Next, we consider the electrochemical potential $\mu$ and use it to derive
Eq.~(\ref{eq:VISS_tanh}). To first order in $T_{xy}$, $\mu$ is spatially
constant in the absence of injected charge currents. At this order, $\mu$
satisfies Eqs.~(\ref{eq:app_mu},
\ref{eq:app_Nochargecurrentsx},
\ref{eq:app_Nochargecurrentsy}),
with the spin accumulation $\mu^s$ evaluated at first order in $T_{xy}$.

The inverse spin--splitter voltage is defined as the voltage drop along the
transverse $x$-direction, averaged over the longitudinal direction $y$,
\begin{equation}
V_{\mathrm{ISS}}
=
\frac{1}{L_y}
\int_{-\frac{L_x}{2}}^{\frac{L_x}{2}}\!dx
\int_{-\frac{L_y}{2}}^{\frac{L_y}{2}}\!dy\;
\partial_x \mu(x,y)\;.
\label{eq:app_VISS_def}
\end{equation}

To obtain a closed expression for $V_{\mathrm{ISS}}$, we exploit current
conservation. In the steady state, the charge current density satisfies
$\partial_k j_k = 0$. Since no charge current enters or leaves the system,
the charge current integrated over the $y$-direction must vanish, yielding
\begin{equation}
0
=
\int_{-\frac{L_y}{2}}^{\frac{L_y}{2}}\!dy\;
\Bigl(
\partial_x \mu
+
T_{xy}\partial_y \mu^s
\Bigr)\;.
\end{equation}

The second term can be evaluated explicitly, leading to
\begin{equation}
\partial_x
\int_{-\frac{L_y}{2}}^{\frac{L_y}{2}}\!dy\;
\mu(x,y)
=
-\,T_{xy}
\bigl[
\mu^s(x,\tfrac{L_y}{2})
-
\mu^s(x,-\tfrac{L_y}{2})
\bigr]\;.
\end{equation}

A further integration over $x$ then yields
\begin{equation}
V_{\mathrm{ISS}}
=
-\,\frac{T_{xy}}{L_y}
\int_{-\frac{L_x}{2}}^{\frac{L_x}{2}}\!dx\;
\bigl[
\mu^s(x,\tfrac{L_y}{2})
-
\mu^s(x,-\tfrac{L_y}{2})
\bigr]\; .
\label{eq:app_VISS_int}
\end{equation}

To proceed, it is convenient to express Eq.~\eqref{eq:app_VISS_int} in terms of
the spin accumulation integrated along the transverse direction. We therefore
define
\begin{equation}
I_x(y)
\equiv
\int_{-\frac{L_x}{2}}^{\frac{L_x}{2}}\!dx\;
\mu^s(x,y)\;.
\label{eq:app_Ix_def}
\end{equation}
In terms of $I_x(y)$, Eq.~\eqref{eq:app_VISS_int} becomes
\begin{equation}
V_{\mathrm{ISS}}
=
-\,\frac{T_{xy}}{L_y}
\left[
I_x\!\left(\tfrac{L_y}{2}\right)
-
I_x\!\left(-\tfrac{L_y}{2}\right)
\right]\;.
\label{eq:app_VISS_Ix}
\end{equation}

The spin accumulation $\mu^s$ obeys the diffusion equation Eq.~\ref{eq:app_mus0}.

Integrating this equation over $x\in[-L_x/2,L_x/2]$ and using the boundary
conditions of vanishing spin current at $x=\pm L_x/2$, one finds that $I_x(y)$
satisfies the effective one--dimensional equation
\begin{equation}
\partial_{yy} I_x(y) = \frac{I_x(y)}{\ell_s^2}\;.
\label{eq:app_Ix_diff}
\end{equation}

The general solution of Eq.~\eqref{eq:app_Ix_diff} reads
\begin{equation}
I_x(y)
=
A \cosh\!\left(\frac{y+\frac{L_y}{2}}{\ell_s}\right)
+
B \sinh\!\left(\frac{y+\frac{L_y}{2}}{\ell_s}\right)\;,
\label{eq:app_Ix_sol}
\end{equation}
where $A$ and $B$ are constants fixed by the boundary conditions in the
$y$-direction. At the lower boundary $y=-L_y/2$, no spin current is injected,
implying
\begin{equation}
\partial_y I_x\!\left(-\tfrac{L_y}{2}\right) = 0\;.
\label{eq:app_Ix_bc1}
\end{equation}
This implies $B = 0$.
At the upper boundary $y=+L_y/2$, the point-like spin injector leads to
\begin{equation}
\partial_y I_x\!\left(\tfrac{L_y}{2}\right)
=
\frac{2G_BW}{\sigma_D}
\bigl[
V_s - \mu^s(0,\tfrac{L_y}{2})
\bigr]\;.
\label{eq:app_Ix_bc2}
\end{equation}
Thus,
\[
A =
\frac{1}{\sinh{\frac{L_y}{\ell_s}}}
\frac{2G_B W \ell_s}{\sigma_D}
\bigl[
V_s - \mu^s(0,\tfrac{L_y}{2})
\bigr]\;.
\]

Using the relation between the injected spin accumulation and its spatial
average, Eq.(\ref{eq:app_mu_avg}),
\begin{equation}
\langle \mu^s \rangle
=
\frac{2G_BW\,\ell_s^2}{\sigma_D L_x L_y}
\bigl[
V_s - \mu^s(0,\tfrac{L_y}{2})
\bigr]\;,
\label{eq:app_mu_avg_final}
\end{equation}
we find
\[
A =
\frac{1}{\sinh{\frac{L_y}{\ell_s}}}
\frac{L_xL_y}{\ell_s}
\langle\mu^s\rangle\;.
\]

Evaluating the solution \eqref{eq:app_Ix_sol} at $y=\pm L_y/2$, one finds
\begin{equation}
I_x\!\left(\tfrac{L_y}{2}\right)
-
I_x\!\left(-\tfrac{L_y}{2}\right)
=
\frac{L_x L_y}{\ell_s}
\langle \mu^s \rangle
\tanh\!\left(\frac{L_y}{2\ell_s}\right)\;.
\end{equation}

Finally substituting this result into Eq.~\eqref{eq:app_VISS_Ix}, we obtain
\begin{equation}
V_{\mathrm{ISS}}
=
-\,T_{xy}\,
\frac{L_x}{\ell_s}\,
\langle \mu^s \rangle\,
\tanh\!\left(\frac{L_y}{2\ell_s}\right)\;,
\label{eq:app_VISS_final}
\end{equation}
which is Eq.~(\ref{eq:VISS_tanh})
in the main text for the inverse spin--splitter
voltage generated by a point-like spin injector.




\bibliography{biblio}

\end{document}